\documentclass[10pt,BP]{iopart}

\usepackage{graphicx}
\usepackage{floatrow}

\begin{document}

\title[Leaf growth is conformal]{Leaf growth is conformal}

\author{Karen Alim}
\address{$^1$ Max Planck Institute for Dynamics and Self-Organization, 37077 G\"ottingen, Germany}
\ead{karen.alim@ds.mpg.de}
\author{Shahaf Armon}
\address{$^2$ Racah Institute of Physics, The Hebrew University of Jerusalem, Jerusalem 91904, Israel}
\author{Boris I.~Shraiman}
\address{$^3$ Kavli Institute for Theoretical Physics, University of California, Santa Barbara, CA, 93106, USA}
\address{$^4$ Department of Physics and Program in Biomolecular Science \& Engineering, University of California, Santa Barbara, CA, 93106, USA}
\author{Arezki Boudaoud}
\address{$^5$ Laboratoire Reproduction et D\'eveloppement des Plantes, Univ. Lyon, ENS de Lyon, CNRS, INRA, F-69364 Lyon Cedex 07, France}

\begin{abstract}
Growth pattern dynamics lie at the heart of morphogenesis. Here, we investigate the growth of plant leaves. We compute the conformal transformation that maps the contour of a leaf at a given stage onto the contour of the same leaf at a later stage. Based on the mapping we predict the local displacement field in the leaf blade and find it to agree with the experimentally measured displacement field to 92\%. This approach is applicable to any two-dimensional system with locally isotropic growth, enabling the deduction of the whole growth field just from observation of the tissue contour.
\end{abstract}

\maketitle
%
Development of organs and organisms generally involves the transformation of simple shapes into more complex ones. Following D'Arcy Thompson, investigating the transformation function that describes how an organ simply morphs over time can identify the direction and magnitude of control at play \cite{Thompson:1942}. Plant leaves are particularly suitable for such an approach as they are mostly planar and at the same time display a huge variety of shapes. 

The mesmerising variety in leaf shape inspired much effort to quantify leaf shape and underlying growth dynamics. Since early measurements to measure leaf growth by tracking ink marks \cite{Avery:1933, Richards:1943}, recently, automated methods to measure the displacement of identifiable features in a leaf have been developed \cite{Ainsworth:2005, Rolland-Lagan:2005, Lee:2006, Wiese:2007, Remmler:2012, Armon:2014}. Quantitative approaches go hand in hand with probing the change in leaf shape resulting from genetic regulation, see Refs.~\cite{Tsukaya:2006, Efroni:2010} for reviews. Yet, in current understanding the underlying biological regulation is too complex to draw definitive conclusions as to what controls growth and thus leaf shape.

Here, the simple geometric transformation morphing the leaf can itself reveal the direction and magnitude of forces at play \cite{Thompson:1942}. For example, a growth field in a linear elastic or viscoelastic material translates to a distributed body force in the equations of equilibrium, with the solutions of the equations of equilibrium determining material shape \cite{Ambrosi:2011}. Constraints on the transformation within this framework entail rules on acting body forces. Can we learn from transformations about the regulation of local cell mechanics?

Here we investigate the transformation mapping successive stages of a growing leaf. We perform the conformal map between the initial and final contour of a planar leaf to predict the displacement field within the leaf. Comparison with experimentally measured displacement fields yields cross-correlations exceeding 92\%. Predicted and measured growth in area both show lowered growth at leaf tip and along the petiole. Growth over the observed time frame from several hours to several days is not a simple dilation, a first order growth, but instead captured by the first two orders of growth; where we identified the orders of growth modes by expanding the predicted conformal map around the leaf base. The surprising fact that a simple conformal map is sufficient to capture the complex dynamics of a growing leaf implies locally isotropic growth.

\section{Materials and Methods.} 
We grew Petunia and Tobacco plants in short-day conditions (8 hours of light per day), and followed leaf shape over time. The leaves chosen were relatively flat and horizontal, and were grown free of any external constrains. A high resolution camera (Luminera Lw575) took top down images of a single leaf every hour. The camera was set on the ceiling in order to minimize growth artifacts due to the leaf movement towards the lens. This artifact ``growth'' scales as: $(L_2-L_1)/L_1 =R/(R-h)-1$ , where $L_1$ is the image size of a leaf placed at distance $R$ from the CCD, and $L_2$ is the image size of the same leaf located at distance $R-h$ from the CCD. By increasing $R$, we lower the effect to less than 0.5\% in area growth. We used a 90mm Tamron lens, and the aperture is set for maximal depth of field (f/32), in order to keep the leaf in focus during growth. A flash of light is scheduled for the picture taking during the night. The measured displacement fields between initial and final frame, $U_x(x,y)$ and $U_y(x,y)$, along the $x$- and the $y$-axis, respectively, are calculated using a particle image velocity algorithm.

The initial and final frame of the picture series are used to extract the leaf's contour line at each time. The conformal map between the initial and final image is calculated with the Schwartz-Christoffel formula \cite{Driscoll:2002} assuming leaf flatness. To account for the rotational degree of freedom three reference points (landmarks) per leaf are placed along the contour line. Including a M\"obius transformation in addition to the conformal mapping gives the resulting map~$f$, that maps the reference points on top of each other.  Reference points for Petunia are leaf tip and sides of the petiole-blade junction; for Tobacco we use white ink marks at the leaf margin applied prior to imaging. Evaluating the conformal map $f$ at every $(x,y)$ position the theoretically calculated displacement fields, $V_x(x,y)$ and $V_y(x,y)$, along the $x$- and the $y$-axis, respectively, are computed. Both measured and calculated displacement fields are corrected for rotation and translation of the entire leaf during growth. Constant angle with respect to the camera is assumed during growth. 
\section{Results and Discussion}
\subsection{Conformal map predicts measured displacement field.}
\begin{figure*}[h!]
\includegraphics[width = \textwidth]{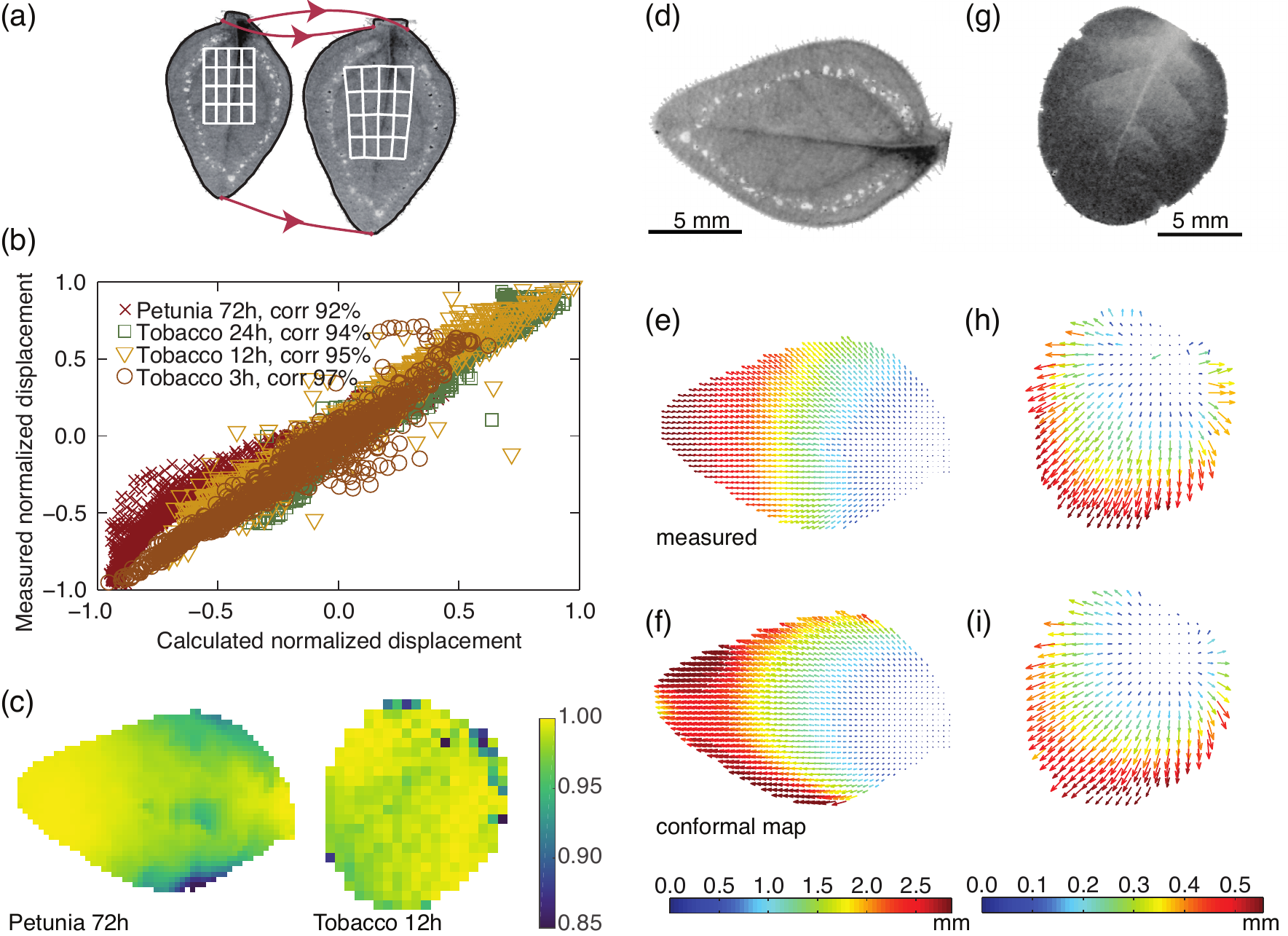}
\caption{\label{fig_panel} Comparison of measured and calculated displacement fields. (a) To calculate the conformal map the contour (\textit{black}) of the initial leaf (\textit{left}) is mapped onto the contour of the final leaf (\textit{right}), rotational degrees of freedom are constrained by specifying the mapping of three reference points (\textit{red}). (b) Correlation between measured and calculated displacement fields at each pixel position in the leaf, normalized by their mean in each spatial component. (c) Correlation by pixel mapped out within two exemplary leaves. (d) Petunia leaf before undergoing 72~h growth and (g) Tobacco leaf before undergoing 12~h growth and their absolute measures of measured (e,h) and calculated (f,i) displacement fields.}
\end{figure*}
We assess the growth of a total of four leaves of two different species with roughly planar leaves, Petunia and Tobacco, over periods as short as three hours and as long as three days. In these time frames specimens grow between 10\% to 42\% in overall size. We compare measured displacement fields with those calculated from a conformal map between the initial and final leaf's contour, see Fig.~\ref{fig_panel} (a). Both measured and calculated displacement fields display the same qualitative characteristics of circular arcs around the leaf tip of ever increasing displacements from leaf base toward leaf tip, see Fig.~\ref{fig_panel} (d-i). To quantify the agreement between measured $U$ and calculated displacement fields $V$ we evaluate the cross-correlation of the vector fields,
\begin{equation}
\textrm{corr} = \frac{\sum_{i\in x,y} (U_i - \langle U\rangle)(V_i - \langle V\rangle)}{\sqrt{\sum_{i\in x,y}(U_i - \langle U\rangle)^2}\sqrt{\sum_{i\in x,y}(V_i - \langle V\rangle)^2}}.
\label{eqn_corr}
\end{equation}
All of our data sets have very high scores of more than 92\% correlation, ranging up to 97\% for the top specimen. Also a pixel-by-pixel comparison of the displacement fields in $x$ and $y$ direction normalised by their mean in $x$ and $y$, respectively, confirm that displacement fields calculated from a conformal map between leaf contours predict the measured displacement fields, see Fig.~\ref{fig_panel}~(b). A detailed mapping of correlation scores per pixel shows that all correlation values are above 85\%, see Fig.~\ref{fig_panel} (c).
\begin{figure}[t]
\includegraphics[width = 8.5cm]{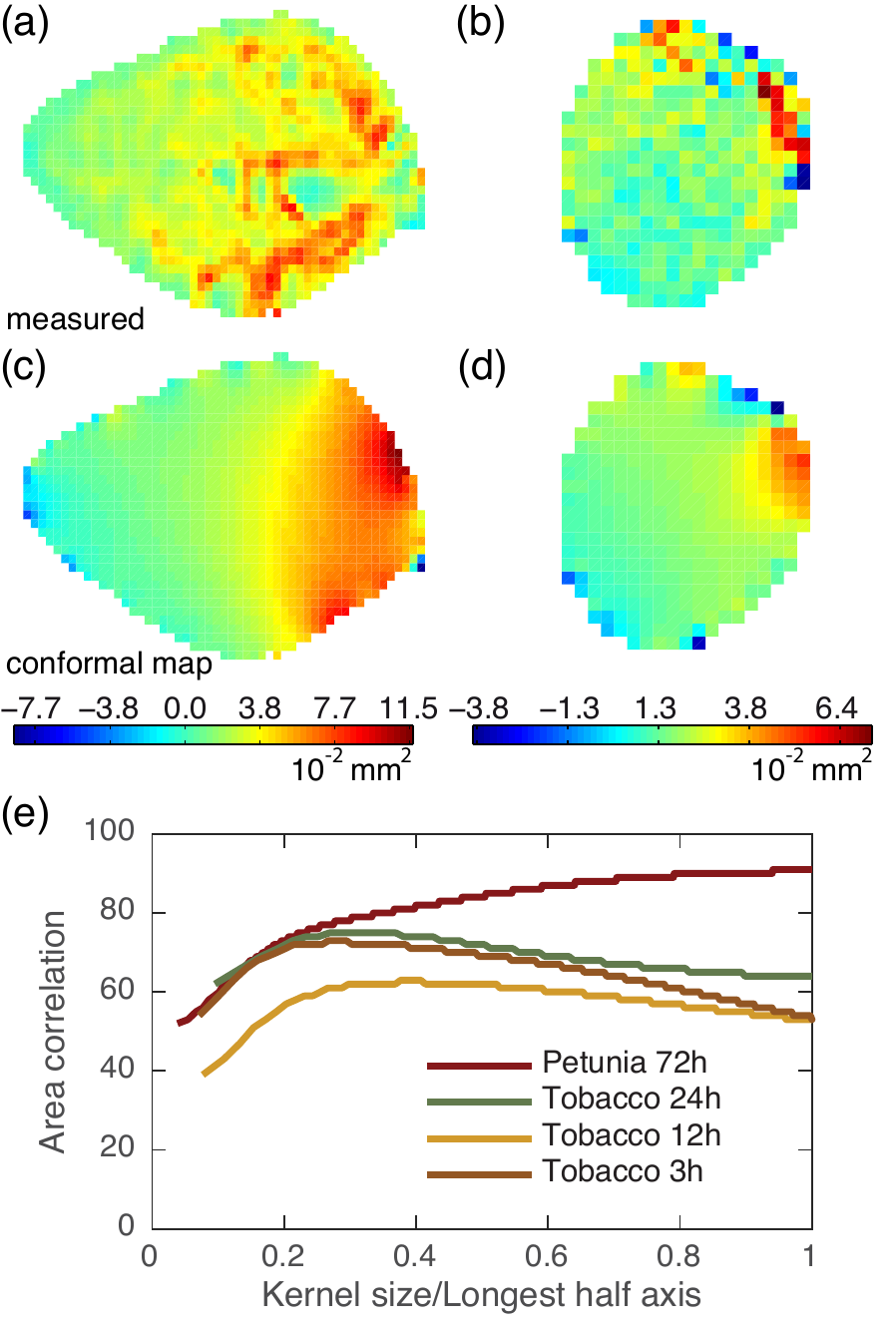}
\caption {\label{fig_growth} (a,b) Measured and (c,d) calculated growth in area of Petunia and Tobacco leaves based on the displacement fields shown in Fig.~\ref{fig_panel} (c-h). Growth is decreasing gradually from leaf base to leaf tip and suppressed along the midvein relative to the leaf blade. (e) Correlation between measured and calculated growth in area as a function of averaging kernel size normalised by each leaf's longest half axis.}
\end{figure}
%

\subsection{Growth fields show characteristic growth dynamics.}
The displacement field is the most direct assessment of leaf growth from both measurements and evaluating the conformal map within the leaf blade, yet, growth in area at each pixel location is the observable that connects to localised growth of cells and cell groups. The individual growth in area $\Delta A$ is given by the divergence of the displacement field at each pixel times pixel area~$\delta A$: $\Delta A=(\frac{\partial U_x}{\partial x}+\frac{\partial U_y}{\partial y})\delta A$, see Fig.~\ref{fig_growth}. Taking the derivative, local fluctuations in growth direction and amplitude diminish the absolute correlations between measured and calculated area growth. Correlation scores in the range 40-60\% are significantly smaller than for the displacement field. If we average out local fluctuations in the measured displacement field the correlation in area growth reaches up to 60-80\% at a kernel size about 1/6 of total leaf size, see Fig.~\ref{fig_growth}(e).

The overall growth patterns are clearly visible in both the measured and the computed growth in area. Growth is largest at the leaf base declining towards the leaf tip in agreement with the growth arrest front that is traveling inward from the leaf tip \cite{Nath:2003}. Growth is reduced along the midvein, as is seen best close to the leaf base. The Petunia leaf also shows small scale patterns in area growth that are not captured by the area growth calculated from the conformal map. Here, growth is reduced along second order veins with interstitial tissue growing faster. 
\begin{figure*}[t]
\includegraphics[width = 0.8\textwidth]{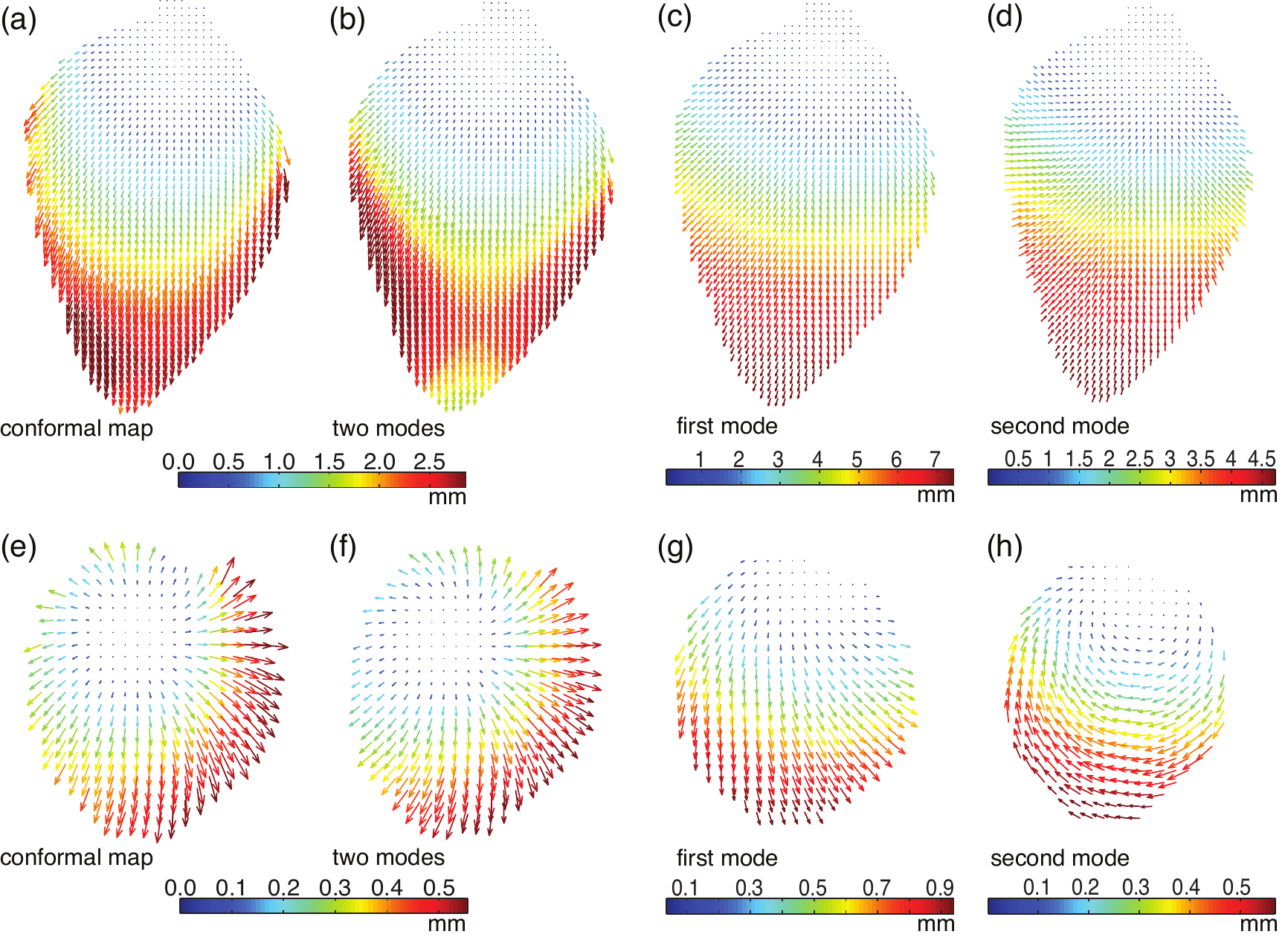}
\caption{\label{fig_expansion} Second order expansion of displacement fields of leaves in Fig.~\ref{fig_panel} around their petiole. Two orders are sufficient to describe the growth in the (a-d) Petunia  and (e-f) Tobacco leaves, although both display substantially different growth dynamics as shown by the decomposition in the first and second mode of growth. Agreement with displacement field arising from the conformal map is above 97\%.}
\end{figure*}

\subsection{Growth dominated by first and second order growth mode in ellipse shaped leaves.}
The conformal map allows us to decompose the overall displacement fields into a series of different modes of growth. To this end we expand the full conformal map 
\begin{equation}
f=\sum_{n=1,2}a_n(z-z_p)^n
\end{equation}
by calculating the best complex Chebyshev approximation of the map $f$~\cite{Fischer:1993} around the position of the petiole~$z_p$, $a_n$ are complex numbers. We neglect higher order modes since their contribution is negligible and the full expansion including the first two orders is already sufficient to reach the same correlation values (see Eq.~\ref{eqn_corr}) as the full conformal map. The first mode alone which is a simple dilation is not sufficient to describe the observed dynamics of growth.  The second mode is necessary to incorporate the low growth rate at the very tip of the leaf. Note that already the combination of two modes of growth allows for a large variety in growth patterns - growth patterns that are captured by a conformal map between initial and final leaf contour.

\begin{figure}[t]
\includegraphics[width = 8.5cm]{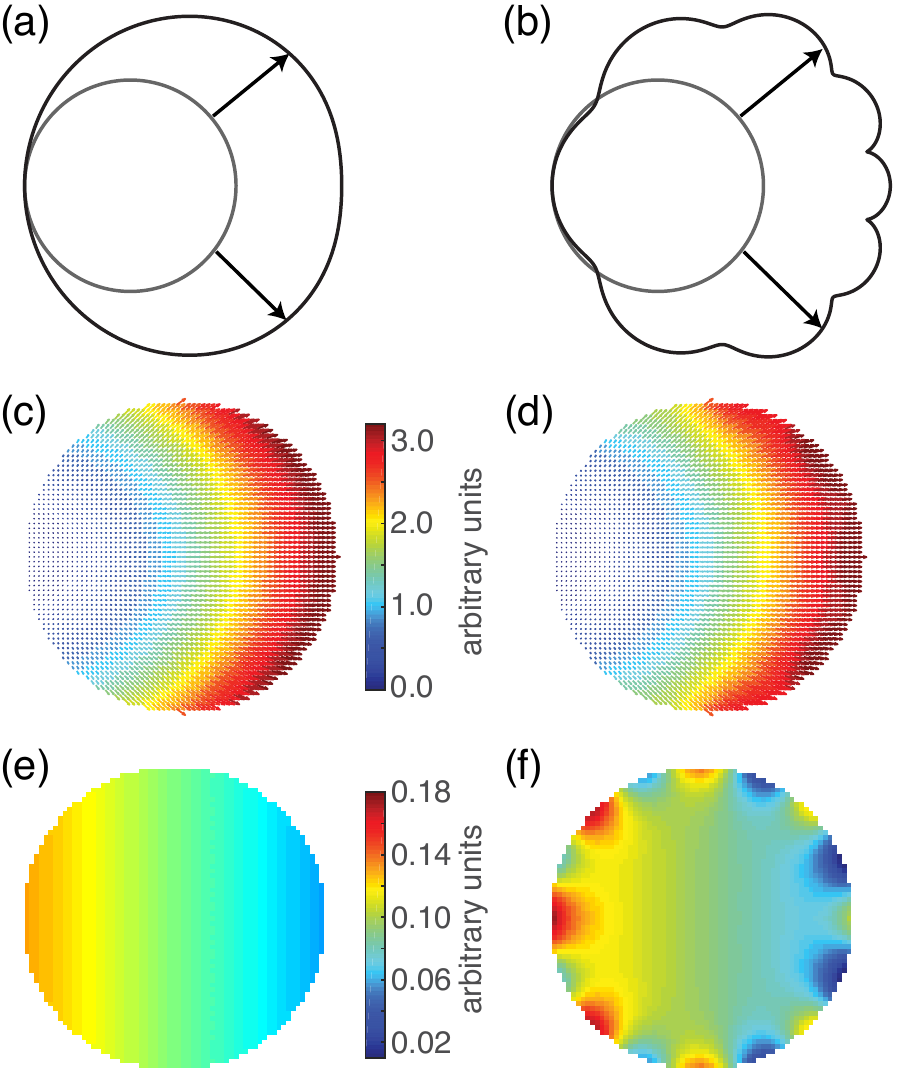}
\caption {\label{fig_theorymap} Higher order growth modes important for growth from simple to multi-lobed leaf shape. Initial ({\it gray}) and final ({\it black}) contour of a theoretical mapping including only the first and second order growth modes (a) and including an additional 9th order growth mode (b). Displacement fields (c,d) and growth in area (e,f) of respective mappings.} 
\end{figure}

To investigate when higher orders become important, we theoretically explored the displacement field and growth in area of a conformal map including one additional higher order term $n>2$ in comparison to a map comprising only first and second order growth mode, see Fig.~\ref{fig_theorymap}. The first two terms give rise to an elliptically shaped contour starting from an initial disc, while the additional higher order term gives rise to a serrated or multi-lobed shape of the final contour. We therefore expect higher order terms to be of general importance when leaves change their shape to higher undulations of their contour during growth.

\subsection{Isotropy and conformal maps.}
We now examine the relation between conformal maps and the strain tensor. If the infinitesimal strain is isotropic, then the diagonal strains are equal, $\partial_x U_x=\partial_y U_y$, while the shear strain vanishes $\partial_y U_x+\partial_x U_y=0$. These two relations are exactly the Cauchy-Schwarz equations that are equivalent to the map being conformal. Conversely, if the map is conformal then strain is isotropic. 

Many plant cells grow anisotropically, which makes it surprising that conformal maps apply to leaves. However in the lateral stages of the growth of lateral plant organs, such as leaves, cotyledons, or sepals, it appears either that cellular growth is roughly isotropic or that growth direction fluctuates spatially and temporally \cite{Zhang:2011,Kuchen:2012,Elsner:2012,Hervieux:2016}. Either case yields isotropic growth at the supracellular level that we are considering here (pixel size $\sim0.6$ mm). Accordingly, we expect our approach to be applicable only to leaves old enough for the blade to have flattened and supracellular growth to have become isotropic.

\section{Conclusion.}
In conclusion we find that a conformal map between initial and final leaf contour predicts both the measured displacement field of a growing leaf and the large scale patterns of growth in area correctly. This is true even though the dynamics of growth of the leaves investigated here are not a mere dilation. We thereby established a tool to study growth dynamics of almost planar two dimensional tissues in a developing organism. This tool is particularly relevant whenever growth is locally isotropic. In future, it would be fascinating to apply this method to a wider range of plant species including leaves with intricate shapes. Here, the conformal map can serve as a baseline to focus particularly on deviations from isotropic growth which might point to mechanical or biochemical constraints of growth like the reduced growth along higher order veins that we observed here.

\section{Acknowledgements}
This research was supported in part by the National Science Foundation under Grant numbers NSF PHY11-25915, NSF PHY05-51164, Deutsche Forschungsgemeinschaft (DFG) via grant SFB-937/A19, European Research Council (Phymorph, StG 307387 to A.B.) and the Deutsche Akademie der Naturforscher Leopoldina (K.A.). 
\\
\providecommand{\newblock}{}

\end{document}